\documentclass[reprint,superscriptaddress,showpacs,nofootinbib,amsmath,amssymb,aps,pra]{revtex4-1}
\usepackage{graphicx}
\usepackage{dcolumn}
\usepackage{bm}
\usepackage[utf8]{inputenc}

\newcommand{\be}{\begin{eqnarray}}
\newcommand{\ee}{\end{eqnarray}}
\newcommand{\nn}{\nonumber}

\begin{document}

\preprint{APS/123-QED}

\title{Magnetoroton in a two-dimensional Bose-Bose mixture}

\author{O. I. Utesov}

\email[Electronic address: ]{utiosov@gmail.com}

\affiliation{Center for Theoretical Physics of Complex Systems, Institute for Basic Science (IBS), Daejeon 34126, Republic of Korea}

\author{S. V. Andreev}
\email[Electronic address: ]{Serguey.Andreev@gmail.com}
\affiliation{Ioffe Institute, 194021 St. Petersburg, Russia}

\date{\today}

\begin{abstract}

We extend our theory of slow magnons in a two-component Bose-Einstein condensate to the case of two spatial dimensions (2D). We provide a detailed discussion of polaronic corrections to the magnon branch of the elementary excitation spectrum in a weakly- and strongly-interacting regimes. In a dilute system, the latter may be achieved by adjusting inter-species attraction such as to obtain a resonance in the $p$-wave scattering channel. Resonantly enhanced $p$-wave attraction results in the self-localization of magnons and formation of a magnetoroton. The nature of a ground state beyond the magnetoroton instability remains to be explored. We suggest a dilute p-wave crystal of alternating polarization as a possible candidate. In contrast to three dimensions (3D), the required strength of the attractive potential in the $s$-wave channel here corresponds to tight binding, and we suggest a potential realization of our model with excitons in 2D semiconductors.

\end{abstract}


\maketitle

\section{Introduction}
\label{Intro}

Quantum degenerate mixtures of bosons have been intensively investigated in a broad context spanning superfluid He$^4$~\cite{Khalatnikov1957, Andreev1975, Mineev1975, Nepomnyashchii1976, Colson1978}, nuclear matter \cite{Migdal1973} and neutron stars~\cite{Babaev2004, Chamel2008, Kobyakov2017}, ultra-cold atomic gases~\cite{Ho1996, Meystre1997, Timmermans1998}, and excitons in semiconductors~\cite{Bobrysheva1988, Andreev2016}. The recent theoretical progress includes a deeper understanding of the superfluid entrainment effect \cite{Fil2005} and its connection to the polaron problem~\cite{Utesov2018, Andreev2020}, phase separation~\cite{Papp2008, Santamore2011, Isaule2021}, elementary excitations~\cite{Utesov2018, Andreev2020, Kim2020} and beyond-mean-field effects~\cite{Petrov2015, Petrov2016, Pastukhov2017, Utesov2018, Arlt2018, Ota2020, Isaule2021}. A mixture of bosons with long-range dipolar forces has been recognized as a promising platform for supersolidity~\cite{Andreev2016, Andreev2017}, which spurred extension of the effective interaction concept along the lines of the Beliaev theory~\cite{Beliaev1958_I, Beliaev1958_II, Utesov2018}.

Particular interest represent polarization waves (``magnons'') produced by oscillations of relative densities of the components. At zero temperature small-amplitude excitations of such kind provide an indicator of miscibility of a Bose-Einstein condensate (BEC) \cite{Mineev1975, Nepomnyashchii1976}. At the verge of the phase separation transition, the magnon sound velocity vanishes and the energy spectrum takes a massive form [\textit{e.g.}, Eq. \eqref{Magnon3D}], reflecting the analogy of an SU(2) symmetric system to a ferromagnet \cite{Halperin1975}. Besides fundamental interest, soft magnons hold practical promise for exciton-mediated superconductivity \cite{Enss2022}.

The simplest model displaying the magnon excitations is a binary mixture of bosons (hereinafter labeled by ``$\uparrow$'' and ``$\downarrow$'') having equal masses $m_\uparrow=m_\downarrow\equiv m$, equal densities $n_\uparrow=n_\downarrow\equiv n$, and elastic two-body interactions which conserve the occupations of each component. The interactions are assumed to produce repulsion in the $s$-wave channels. In a three-dimensional space (3D), analysis of such a model by using the Beliaev approach \cite{Beliaev1958_I, Beliaev1958_II} has revealed a previously unknown forward-scattering correction to the magnon dispersion \cite{Utesov2018}
\begin{equation}
\label{Magnon3D}
\hbar\omega(\bm p)=\frac{\hbar^2 p^2}{2m}+2n[g\left(\tfrac{\bm p}{2},\tfrac{\bm p}{2}\right)-g(0,0)],
\end{equation}
where the energy-dependent effective interaction potential $g(\bm k, \bm k)$ is defined by the forward (zero-angle) scattering amplitude as $g(\bm k, \bm k)=-(4\pi\hbar^2/m)\mathrm{Re}[f(\bm k, \bm k)]$. The massive form of Eq. \eqref{Magnon3D} corresponds to the particular case of identical interactions. The necessity to retain only the real part of the amplitude will be shown in Section \ref{Generalities} of the present paper.

The forward-scattering amplitude is known to govern refraction of matter waves \cite{LandauQM, Tannoudji2011}. A plane wave $\sim e^{i\bm k\bm r - i\omega t}$ propagating through a homogeneous medium of fixed scatterers distributed at the density $2n$ undergoes a shift of its kinetic energy $\hbar^2k^2/2m=\hbar^2 \bar k^2/2m+(4\pi\hbar^2n/m) f(\bar{\bm k}, \bar{\bm k})$ \cite{LandauQM}. Here $\bar{\bm k}$ is the momentum in vacuum, its absolute value being fixed by the total (constant) energy $\hbar\omega$ through the dispersion relation $\bar k=\sqrt{2m \omega/\hbar}$. By writing $\bm k=\mathsf{n}(\omega) \bar{\bm k}$, one may see that the quantity $\mathsf{n}(\omega)\equiv \sqrt{1+(2\pi\hbar n/m \omega) f(\bar{\bm k}, \bar{\bm k})}$ plays the role of refraction index. The imaginary part of $\mathsf{n}(\omega)$ defines attenuation of the particle beam, in agreement with the optical theorem \cite{LandauQM}. The in-medium dispersion becomes $\hbar\omega(\bm k)=(\hbar^2k^2/2m)\mathsf{n}^2(\omega)$. By counting the energy from the mean-field potential $2ng(0,0)$, replacing the incident particle mass $m$ and momentum $\bm k$ by the reduced mass $m/2$ and the momentum $\bm p/2$ relative to the condensate, respectively, and \textit{suppressing the attenuation}, one gets Eq.~\eqref{Magnon3D}.

Alternatively, one may regard the forward-scattering term in Eq. \eqref{Magnon3D} as a correction to the quasiparticle mass \cite{Utesov2018, Andreev2020}. This viewpoint has been corroborated by the diagrammatic theory \cite{Utesov2018} revealing renormalization of the magnon mass due to the Andreev-Bashkin entrainment effect \cite{Andreev1975, Fil2005} and establishing a connection to the physics of a Bose polaron \cite{Bruun2015}. The polaronic interpretation further suggests an effect beyond the optical analogy: \textit{self-localization} of magnons \cite{Andreev2020}. This corresponds to the strong coupling regime in an analogous impurity problem, where the particle sticks to the medium rather than drags a virtual cloud. A challenging question is how to achieve the strong coupling in a weakly-interacting gas $n |f^3(0,0)|\ll 1$, so that the quantum depletion of BEC remains small.

To this end, one may notice \cite{Andreev2020}, that for distinguishable bosons the forward-scattering amplitude $f(\bm k,\bm k)$ may include \textit{odd} partial waves, of which the leading contribution at low energies would be due to $p$-waves. More generally, if one moves away from the phase separation point into the miscibility region $g_{\uparrow\downarrow}(0,\bm p)<g_{\uparrow\uparrow}(0,\bm p)$ by allowing a difference between intra- and inter-species interactions, $g_{\uparrow\uparrow}(\bm p,\bm q)= g_{\downarrow\downarrow}(\bm p,\bm q)$ and $g_{\uparrow\downarrow}(\bm p,\bm q)$, respectively, the magnon spectrum takes the form \cite{Andreev2020}
\begin{equation}
\label{SpinWave3D}
\hbar\omega(\bm p)=\sqrt{\frac{\hbar^2p^2}{2m_\ast}\left[\frac{\hbar^2p^2}{2m_\ast}+2n[g_{\uparrow\uparrow}(0,\bm p)-g_{\uparrow\downarrow}(0,\bm p)]\right]},
\end{equation}
where the renormalized mass $m_\ast$ is defined through the identity
\begin{widetext}
\begin{equation}
\label{MagnonMass}
\frac{\hbar^2p^2}{2m_\ast}\equiv\frac{\hbar^2p^2}{2m}+2n\left[g_{\uparrow\downarrow}^{-}\left(\tfrac{\bm p}{2},\tfrac{\bm p}{2}\right)+g_{\uparrow\uparrow}^{+}\left(\tfrac{\bm p}{2},\tfrac{\bm p}{2}\right)-g_{\uparrow\uparrow}(0,0)\right]+n\left[\delta g_{\uparrow\downarrow}(0,\bm p)-\delta g_{\uparrow\uparrow}(0,\bm p)\right]
\end{equation}
\end{widetext}
with
\begin{subequations}
\begin{align}
g_{\sigma\sigma'}^{\pm}(\bm k,\bm k)&\equiv\tfrac{1}{2}[g_{\sigma\sigma'}(\bm k,\bm k)\pm g_{\sigma\sigma'}(\bm k,-\bm k)]\\
\label{OffShellCorrections}
\delta g_{\sigma\sigma'}(0,\bm k)&\equiv g_{\sigma\sigma'}(0,\bm k)-g_{\sigma\sigma'}(0,0).
\end{align}
\end{subequations}
The formula \eqref{Magnon3D} is obtained from Eq. \eqref{SpinWave3D} in the limit $g_{\uparrow\uparrow}(\bm p,\bm q)= g_{\downarrow\downarrow}(\bm p,\bm q)=g_{\uparrow\downarrow}(\bm p,\bm q)\equiv g(\bm p,\bm q)$.

The off-shell corrections \eqref{OffShellCorrections} vanish at low momenta, provided the bare microscopic interactions are short-ranged. Likewise, the product  $n[g_{\uparrow\uparrow}^{+}(\bm k,\bm k)-g_{\uparrow\uparrow}(0,0)]$ may be omitted in the weakly-interacting limit $n f^3(0,0)\ll 1$. One may then see, that the prerogative is left entirely to the antisymmetrized potential $g_{\uparrow\downarrow}^{-}(\bm k,\bm k)$, governed by the $p$-wave scattering. This is in stark contrast to the phonon part of the spectrum, where the inter-species interaction enters in the symmetrized form $g_{\uparrow\downarrow}^{+}(\bm k,\bm k)$, and, therefore, only even partial wave channels are relevant \cite{Andreev2020}.

On the basis of Eq. \eqref{MagnonMass} we have shown \cite{Andreev2020}, that an increasingly strong enhancement of the magnon mass $m_\ast$ may be achieved on the attractive side of a $p$-wave resonance, at the same time retaining low level of quantum fluctuations due to phonons. Moreover, since the bare magnons with energies $E$ closer to the resonance stick stronger to the medium, self-localization may be achieved at finite momentum $\bm p\neq 0$. The corresponding dip in the magnon dispersion has been dubbed \textit{magnetoroton} and has been tentatively associated with an unconventional phase separation transition \cite{Andreev2020}, which yet remains to be explored. Quite distinctly from the conventional softening of the magnon mode at the verge of miscibility, self-localization occurs upon increasing the particle density $n$ while keeping the ratio $g_{\uparrow\downarrow}(0,0)/g_{\uparrow\uparrow}(0,0)<1$ fixed.

However, experimental implementation of these ideas in 3D has been hindered by difficult access to a $p$-wave resonance. A centrifugal $p$-wave resonance naturally emerges from an $s$-wave bound state when the energy of the latter is pushed toward the scattering threshold. In 3D the corresponding binding energy corresponds to the limit of weak binding \cite{Andreev2020}, which is rather exotic. In principle, a $p$-wave resonance exists as the Feshbach resonance in mixtures of ultra-cold atomic gases, such as $^{85}$Rb-$^{87}$Rb mixtures \cite{Bohn2004, Papp2008}. Unfortunately, the three-body loss of atoms near the resonance has been detrimental \cite{Levinsen2008}.

In this regard, the reduction of spatial dimensionality has been considered a promising route for the realization of $p$-wave attraction. Indeed, as we argue in this paper, a centrifugal $p$-wave resonance in two dimensions (2D) would originate from a tightly bound $s$-wave state. Such states exist for excitons in 2D semiconductors, and we discuss a possible route in this direction. Importantly, both the reduced spatial dimensionality and the large binding energy slow down the relaxation of a mixture to the $s$-wave molecular condensate. In atomic settings, one may take advantage of the residual zero-point motion in the transverse direction to increase the ratio of equilibration to three-body recombination rates \cite{Levinsen2008}. Besides these specific advantages, the 2D geometry is generically more convenient for the eventual quest of a new ground state (both in theory and experiment).

Technically, an extension of the formal result \eqref{SpinWave3D} to 2D is far from being trivial. Preliminary considerations \cite{Utesov2018} suggest that the concept of effective interaction does not apply literally in this case. Enhanced quantum fluctuations and, concomitantly, stronger superfluid drag look promising, but do not imply self-localization which would require the contribution of the forward scattering. In fact, 2D quantum kinematics is quite distinct from 3D in that cold particles tend to avoid each other at short distances \cite{LandauQM}. Hence, the possible existence of magnetorotons in 2D yet remains to be proven at the conceptual level. Furthermore, in contrast to 3D, short-range forces are no longer ubiquitous in 2D, where interacting BECs have been realized primarily with dipolar excitons in semiconductor layers \cite{Butov2016}. Long-range dipolar repulsion is known to induce sizeable momentum dependence of the off-shell scattering amplitude \cite{Shlyapnikov2011, Baranov2011, Shlyapnikov2013, Andreev2015, Andreev2016}, so that corrections of the type \eqref{OffShellCorrections} may come into play. These open questions have provided the main motivation for an extension of the magnetoroton theory presented in this paper.

Our main result is that close to the $p$-wave resonance Eq. \eqref{SpinWave3D} remains valid, whereas the expression for the effective mass \eqref{MagnonMass} reduces to Eq. \eqref{MagnonMass2D_generic} governed entirely by the short-range part of the $p$-wave scattering amplitude. The long-range parts cancel identically and do not contribute to the mass renormalization. The above qualitative description of the magnetoroton in 3D pertains also to the 2D space. Crucially, only the real part of the scattering amplitude contributes to the effective potential in Eq. \eqref{MagnonMass2D_generic}: the condensate stabilizes the magnetoroton. Furthermore, we find that the magnetoroton is the only possible form of the magnon self-localization in a 2D gas, \textit{i.e.}, the scenario with $m_\ast\rightarrow \infty$ at $\bm p\rightarrow 0$ is absent here.

To reach these conclusions, we start with a general analysis of the polaronic corrections to the magnon dispersion based on the diagrammatic approach of Ref. \cite{Utesov2018} and quantum scattering theory. This analysis is presented in Section~\ref{Generalities}. In Section~\ref{Implementation} we discuss the experimental implementation of the magnetoroton with dipolar bosons in bilayers. We predict a $p$-wave scattering resonance in the limit where the inter-layer potential supports a tightly-bound $s$-wave state. At low temperatures, relaxation into an $s$-wave molecular condensate is inhibited by several factors, including the long-range dipolar repulsion. This warrants the existence of a mixture as a metastable state on the attractive side of the $p$-wave resonance and consecutive dynamical instability upon softening the magnetoroton. The nature of a new phase-separated ground state remains to be explored. We tentatively suggest dipolar excitons in atomically thin semiconductors for an immediate experimental test of our findings.

\section{General consideration}

\label{Generalities}

In the frame of the Beliaev approach, the elementary excitation spectrum of a Bose-Einstein condensate is obtained from the poles of the one-particle Green's function \cite{Beliaev1958_I, Beliaev1958_II, Schick1971, Lozovik1978}. For a two-component condensate the spectrum $\hbar\omega(\bm p)$ has two branches that correspond to the in-phase and out-of-phase oscillations of the components - the phonon and magnon elementary excitations, respectively \cite{Utesov2018}. Each branch is formally expressed in terms of the (bare) single-particle kinetic energy, the chemical potential $\mu$ and the self-energies $\Sigma^{\sigma\sigma^\prime}_{11}(\pm\mathsf p)$ and $\Sigma^{\sigma\sigma^\prime}_{20}(\mathsf p)$ with $\mathsf p\equiv (\omega, \bm p)$ being the $4$-momentum which itself includes the excitation energy $\hbar\omega (\bm p)$. The definition and physical meaning of the three quantities $\Sigma^{\sigma\sigma^\prime}_{11}(\pm\mathsf p)$ and $\Sigma^{\sigma\sigma^\prime}_{20}(\mathsf p)$ can be found in our earlier work \cite{Utesov2018}. Here, we focus exclusively on the magnon branch [Eq. (26) in Ref. \cite{Utesov2018} with the sign-choice ``$-$''], and assume that the translational motion and scattering of particles is reduced to two dimensions, \textit{i.e.} $\bm p=(p_x,p_y)$.

The basic element of perturbative expansion of the self-energies is the four-leg vertex $\Gamma_{\sigma\sigma^\prime\sigma\sigma^\prime} (\mathsf p_1,\mathsf p_2; \mathsf p_3,\mathsf p_4)$ which represents an infinite series of the ladder diagrams for the bare two-body interaction potentials $V_{\sigma\sigma^\prime}(r)$. The vertex may be related to the two-body scattering matrix $T_{\sigma\sigma^\prime}(\bm k^\prime,\bm k, z)$ by putting $z=\Omega-P^2/4m+2\mu+i 0$ with $(\Omega,\bm P)\equiv \mathsf P=\mathsf p_1+\mathsf p_2=\mathsf p_3+\mathsf p_4$ being the total energy and momentum of the colliding particles, $\bm k=(\bm k_1-\bm k_2)/2$ and $\bm k^\prime=(\bm k_3-\bm k_4)/2$ being the corresponding relative momenta. The Bethe-Salpeter equation for the vertex translates into the following integral equation for the scattering matrix:
\begin{multline}\label{Vert2}
T_{\sigma\sigma^\prime}(\bm k^\prime,\bm k; z) = \frac{1}{(2\pi)^2}V_{\sigma\sigma^\prime}(\textbf{k}^\prime-\bm k)\\
 +\frac{1}{(2\pi)^2} \int \frac{V_{\sigma\sigma^\prime}(\textbf{k}^\prime - \textbf{k}^{\prime\prime})}{z-E_{\bm k^{\prime\prime}}}T_{\sigma\sigma^\prime}(\bm k^{\prime\prime},\bm k; z) d\bm k^{\prime\prime},
\end{multline}
where $E_{\bm k}=\hbar^2 k^{2}/m$ and the Fourier transform of the bare interaction potentials is defined as
\begin{equation}
V_{\sigma\sigma'}(\bm q)=\int e^{-i\bm q\bm x}V_{\sigma\sigma'}(\bm x)d\bm x.
\end{equation}
The $T$-matrix is related to the off-shell scattering amplitude
\begin{equation}
\label{ScatteringAmplitude}
f_{\sigma\sigma^\prime}(\bm k^\prime,\bm k)\equiv-(2\pi)^2\frac{m}{2\hbar^2}T_{\sigma\sigma^\prime}(\bm k^\prime,\bm k; E_{\bm k}+i0)
\end{equation}
by
\begin{widetext}
\begin{equation}
\label{Vert3}
T_{\sigma\sigma^\prime}(\bm k^\prime,\bm k; z)=-\frac{1}{(2\pi)^2}\frac{2\hbar^2}{m}\Bigl [f_{\sigma\sigma^\prime}^{\ast}(\bm k,\bm k^\prime)-
\frac{1}{(2\pi)^2}\frac{2\hbar^2}{m}\int f_{\sigma\sigma^\prime}(\bm k^\prime,\bm q)f_{\sigma\sigma^\prime}^{\ast}(\bm k,\bm q)\left (\frac{1}{E_{\bm q}-E_{\bm k^\prime}+i0}+\frac{1}{z-E_{\bm q}}\right)d\bm q\Bigr].
\end{equation}
\end{widetext}
To the first order in the dimensionless parameters
\begin{equation}
\label{WeakInt}
\eta_{\sigma\sigma^\prime}\equiv \frac{m}{\hbar^2}(2\pi)^2 T_{\sigma\sigma^\prime}(0,0; 2\mu+i0)\ll 1
\end{equation}
the self-energies and the chemical potential may be expressed as
\begin{equation}
\label{SigmaViaTs}
\begin{split}
\Sigma^{\sigma\sigma^\prime}_{11}(\pm\mathsf p)&=(2\pi)^2 n [T_{\sigma\sigma^\prime}(\mp\bm p/2,\pm\bm p/2;z_\pm)\\
&+\delta_{\sigma\sigma^\prime}\sum_{\sigma^{\prime\prime}}T_{\sigma\sigma^{\prime\prime}}(\pm\bm p/2,\pm\bm p/2;z_\pm)]\\
\Sigma^{\sigma\sigma^\prime}_{20}(\mathsf p)&=(2\pi)^2 n T_{\sigma\sigma^\prime}(0,\bm p;2\mu+i0),
\end{split}
\end{equation}
with
\begin{equation}
\label{EnergeticArgument}
z_\pm=\pm\hbar\omega-
\frac{\hbar^2 p^2}{4m}+2\mu+ i0
\end{equation}
and
\begin{equation}
\label{mu}
\mu=(2\pi)^2 n[T_{\uparrow\uparrow}(0,0; 2\mu+i0)+T_{\uparrow\downarrow}(0,0; 2\mu+i0)],
\end{equation}
respectively. By using the well-known solution of a 2D scattering problem in the $s$-wave channel
\begin{equation}
\label{Swave}
f_{0,\sigma\sigma^\prime}(k)=\frac{\pi}{\ln(k a_{\sigma\sigma^\prime})-i\pi/2}
\end{equation}
and the relation \eqref{Vert3}, the transcendental equation for the chemical potential may also be recast as
\begin{equation}
\label{mu2D}
\mu=-\frac{\hbar^2 n}{m}\left [\frac{2\pi}{\ln (p_c a_{\uparrow\uparrow})}+\frac{2\pi}{\ln (p_c a_{\uparrow\downarrow})}\right],
\end{equation}
where we have defined $p_c\equiv\sqrt{2m\mu}/\hbar$. The requirement $\eta_{\sigma\sigma^\prime}\ll 1$ is then identical to either $p_c a_{\sigma\sigma^\prime}\ll 1$ or $p_c a_{\sigma\sigma^\prime}\gg 1$. Here, the 2D scattering length $a_{\sigma\sigma^\prime}$ would be on the order of the microscopic range $R_{\sigma\sigma^\prime}$ of the bare potential $V_{\sigma\sigma'}(r)$, in the case where the latter possesses a hard core at $r\lesssim R_{\sigma\sigma^\prime}$. In the opposite limit of a soft-core potential $m |V_{\sigma\sigma^\prime}(\bm q)|/\hbar^2\ll 1$, one would have instead $a_{\sigma\sigma^\prime}\sim R_{\sigma\sigma^\prime} \exp[-4\pi\hbar^2/m V_{\sigma\sigma^\prime}(0)]$, which yields the Born approximation for the scattering amplitude. Interestingly, in that latter limit, the diluteness criterion $\beta_{\sigma\sigma^\prime}\equiv\sqrt{nR_{\sigma\sigma^\prime}^2}\ll 1$ may be to a certain extent relaxed so that the theory would be applicable also at moderate densities $\beta_{\sigma\sigma^\prime}\lesssim 1$ \cite{Lozovik1978}.

Hereinafter, we shall be concerned with the repulsive Bose-Bose mixtures, where $p_c a_{\sigma\sigma^\prime}\ll 1$. The miscibility condition $T_{\uparrow\downarrow}(0,0; 2\mu+i0)<T_{\uparrow\uparrow}(0,0; 2\mu+i0)$ may be recast as $a_{\uparrow\downarrow}<a_{\uparrow\uparrow}$. At $a_{\uparrow\downarrow}>a_{\uparrow\uparrow}$ phase separation of the system and formation of immiscible domains would occur.

As we have outlined above, the elementary excitation spectrum requires calculation of the momentum-dependent self-energies \eqref{SigmaViaTs}. There may be two qualitatively distinct contributions to these relations. The first one comes from the energetic arguments of the $T$-matrices given by Eq. \eqref{EnergeticArgument}. This contribution may be analyzed by setting $\hbar\omega\approx \hbar c_{m} p$ (or $\hbar\omega\approx \hbar^2 p^2/2m$ at the verge of miscibility $a_{\uparrow\downarrow}=a_{\uparrow\uparrow}$) and expanding the $T$-matrices in powers of $p/p_c$ by using Eq. \eqref{Vert2}. The first-order correction goes as the polaronic renormalization of the magnon mass $m$ (or, equivalently, the spin-wave velocity $c_m$) and may be identified with the superfluid drag $\rho_{\uparrow\downarrow}$ due to the Andreev-Bashkin entrainment effect \cite{Utesov2018, Konietin2018}. This contribution scales as the quantum depletion of the condensate $\rho_{\uparrow\downarrow}\sim n^\prime=n-n_0\sim n\eta_{\sigma\sigma^\prime}$ and may safely be neglected in our consideration. \footnote{It would, on the other hand, be important for quantum-mechanical stabilization of a collapsing Bose-Bose mixture with interspecies attraction \cite{Petrov2016}.}

The second contribution comes from the explicit momentum dependence of the $T$-matrices. For the long-range potentials, this includes the so-called anomalous scattering \cite{LandauQM}. The anomalous momentum-dependent correction to the off-shell scattering amplitude \eqref{ScatteringAmplitude} may be worked out in the Born approximation. The spectrum then takes the form of Eq. \eqref{SpinWave3D}, where
\begin{equation}
\label{EffPotentials}
g_{\sigma\sigma^\prime}(\bm p, \bm q)\equiv (2\pi)^2T_{\sigma\sigma^\prime}(\bm p, \bm q, 2\mu+i0)
\end{equation}
are the properly defined effective potentials. Remarkably, Eq. \eqref{SpinWave3D} may be obtained by applying the Bogoliubov transformation to the second-quantized Hamiltonian
\begin{widetext}
\begin{equation}
\label{DiluteHamiltonian}
\hat H_\ast=\sum_{\textbf{p},\sigma}\frac{\hbar^2 p^2}{2m}\hat a_{\sigma, \textbf{p}}^{\dag} \hat a_{\sigma, \textbf{p}}+\frac{1}{2S}\sum_{\textbf k,\textbf p,\textbf q,\sigma,\sigma^\prime}\hat a_{\sigma, \textbf k+\textbf p}^{\dag} \hat a_{\sigma^\prime,\textbf k-\textbf p}^{\dag} g_{\sigma\sigma^\prime}(\bm p, \bm q)\hat a_{\sigma, \bm k+\bm q}\hat a_{\sigma^\prime,\bm k-\bm q},
\end{equation}
\end{widetext}
where $S$ is the quantization area and the operators $\hat a_{\sigma, \textbf{p}}$ obey the standard Bose commutation relations. The transcendental equation \eqref{mu} for the chemical potential may be obtained in the zeroth order of the perturbative expansion of $\hat H_\ast$ upon replacement of $\hat a_{\sigma, 0}$'s by the $c$-numbers $\sqrt{N}$.

Let us single out the anomalous contributions to the relevant potentials:
\begin{subequations}
\begin{align}
g^{\pm}_{\sigma\sigma^\prime}(\tfrac{\bm p}{2},\tfrac{\bm p}{2})&\equiv g^{(\pm)}_{\sigma\sigma^\prime}(\tfrac{p}{2},\tfrac{p}{2})\pm \tfrac{1}{2}g^{\ast}_{\sigma\sigma^\prime}(p)\\
g_{\sigma\sigma^\prime}(0, \bm p)&\equiv g_{\sigma\sigma^\prime}(0,p)+g^{\ast}_{\sigma\sigma^\prime}(p).
\end{align}
\end{subequations}
Here, the normal symmetrized (antisymmetrized) contributions $g^{(\pm)}_{\sigma\sigma^\prime}(\tfrac{p}{2},\tfrac{p}{2})$ correspond to the even (odd) partial series in the multipolar expansion of the on-shell scattering amplitude
\begin{equation}
\label{MultipolarSeries}
f_{\sigma\sigma^\prime}(\bm k^\prime,\bm k)=\sum\limits_{m=0}^{+\infty}f_{m,\sigma\sigma^\prime}(k)e^{im\varphi}
\end{equation}
with $\varphi$ being the angle between $\bm k^\prime$ and $\bm k$. The $m=0$ term in Eq. \eqref{MultipolarSeries} is just the $s$-wave scattering amplitude given by Eq. \eqref{Swave}. The anomalous contributions $g^{\ast}_{\sigma\sigma^\prime}(p)$ can be seen to cancel exactly in Eq. \eqref{MagnonMass}. On the other hand, $g^{\ast}_{\sigma\sigma^\prime}(p)$'s may survive in the effective mean-field potential produced by the condensate in Eq. \eqref{SpinWave3D}. As such, these contributions may give rise to the so-called roton immiscibility \cite{Wilson2012} - a phase separation at finite momentum driven by minimization of the interaction energy. This phenomenon has been extensively addressed elsewhere (see Ref. \cite{Andreev2020} and references therein) and will be intentionally eliminated from our present consideration by assuming $g^{\ast}_{\uparrow\uparrow}(p)=g^{\ast}_{\uparrow\downarrow}(p)$.

Furthermore, at low $p$ the leading $s$-wave contributions to the ordinary parts $g^{(+)}_{\uparrow\uparrow}(\tfrac{p}{2},\tfrac{p}{2})$, $g_{\uparrow\uparrow}(0,p)$ and $g_{\uparrow\downarrow}(0,p)$ cancel identically with $g_{\uparrow\uparrow}(0,0)$ and $g_{\uparrow\downarrow}(0,0)$, respectively. This is consistent with the absence of a gap for a Goldstone mode. One obtains
\begin{equation}
\label{MagnonMass2D_generic}
\frac{\hbar^2p^2}{2m_\ast}=\frac{\hbar^2p^2}{2m}+2n g^{(-)}_{\uparrow\downarrow}(\tfrac{p}{2},\tfrac{p}{2}),
\end{equation}
where $g^{(-)}_{\uparrow\downarrow}(\tfrac{p}{2},\tfrac{p}{2})$ is governed entirely by the $p$-wave scattering amplitude
\begin{equation}
\label{PwaveAmplitude}
f_{1,\uparrow\downarrow}(k)=\frac{\pi}{-\tfrac{E_{\bm k}-\nu}{\beta E_{\bm k}}+\ln(k R_{\uparrow\downarrow})-i\pi/2}
\end{equation}
through the relations \eqref{EffPotentials} and \eqref{Vert3}. Here, as we have already defined above, $E_{\bm k}\equiv\hbar^2 k^2/m$. At $k\rightarrow 0$ one may introduce the $p$-wave scattering area $s\equiv -\beta \pi \hbar^2/(m\nu)$ and get $m_\ast=m/(1-2n|s|)$. As the density $n$ approaches the critical value
\begin{equation*}
n_c^{(1)}=(2|s|)^{-1},
\end{equation*}
the effective mass $m_\ast$ diverges.

By analogy with the 3D case considered in our previous work \cite{Andreev2020}, one would expect that the divergency of $m_\ast$ becomes compatible with the diluteness criterion when one approaches a $p$-wave scattering resonance from below. The parameter $0<\nu\ll \hbar^2/mR_{\uparrow\downarrow}^2$, in this case, would play the role of ``detuning'' and $\beta$ would characterize the width of the resonance. Thus, for a rectangular potential well of depth $U_{\uparrow\downarrow}$ and radius $R_{\uparrow\downarrow}$, one would have $\beta = 1 / \ln{(2 e^{1/2 -\gamma})}$ with $\gamma$ being the Euler-Macheroni constant, and $\nu=\beta (U_{\uparrow\downarrow}^{(c)}-U_{\uparrow\downarrow})/2$ with $U_{\uparrow\downarrow}^{(c)}\approx 5.76\hbar^2/mR_{\uparrow\downarrow}^2$ being the threshold depth at which a $p$-wave resonance emerges (see Appendix \ref{RectangularWell}).

At this point, it is instructive to return to our argument on the refractive origin of the polaronic effect (Section \ref{Intro}). In a hypothetical medium of rigid scatterers, the large imaginary part of the scattering amplitude near a resonance would inevitably produce strong attenuation of the matter wave, just like divergency of the conventional refractive index signals absorption of light. Crucially, the attenuation does not occur for magnons. A formal proof of this statement may be obtained from Eq. \eqref{Vert3}. In the relevant limit \eqref{WeakInt} the only significant contribution to the imaginary part of the integral term is due to the unbound denominator $(E_{\bm q}-E_{\bm k^\prime}+i0)^{-1}$. By virtue of the optical theorem \cite{LandauQM}
\begin{equation}
\begin{split}
&\mathrm{Im}[f_{\sigma\sigma^\prime}(\pm\bm k,\bm k)]=\\
&\frac{1}{4\pi}\int d\bm q f_{\sigma\sigma^\prime}(\bm k,\bm q)f_{\sigma\sigma^\prime}^\ast(\pm\bm k,\bm q)\delta(q^2-k^2),
\end{split}
\end{equation}
this contribution cancels identically with the imaginary part of $f_{\sigma\sigma^\prime}(\pm\bm k,\bm k)$. One gets
\begin{equation}
g^{(-)}_{\uparrow\downarrow}(\tfrac{p}{2},\tfrac{p}{2})=-\frac{2\hbar^2}{m}\mathrm{Re}[f_{1,\uparrow\downarrow}(\tfrac{p}{2})].
\end{equation}
Qualitatively, attenuation of a matter wave near a scattering resonance would mean capture by the scatterers. For magnons, this would correspond to conversion into $p$-wave molecules, and in a condensate the latter may occur only as a phase transition.

This way, the condensate stabilizes the magnons and endows them with the possibility to explore the whole resonant pole structure of Eq. \eqref{PwaveAmplitude}. The most dramatic modification of the dispersion law takes place at small detuning $\nu$. The magnons with their energies being close to the resonance exhibit stronger renormalization of their mass so that the spectrum \eqref{SpinWave3D} develops a minimum at finite momentum: \textit{magnetoroton}. At the critical density
\begin{equation}
\label{SecondDensity}
n_c^{(2)}\approx - \dfrac{m\nu}{\beta \hbar^2\ln(2 m R_{\uparrow\downarrow}^2 \nu/ \beta \hbar^2)}
\end{equation}
the minimum touches zero and the mixture becomes dynamically unstable. One may see that, in contrast to 3D \cite{Andreev2020}, here one has $n_c^{(2)}\ll n_c^{(1)}$ due to the large logarithm in the denominator. Hence, self-localization of magnons in a dilute 2D gas may occur only via the magnetoroton. The characteristic momentum of the instability (the magnetoroton position) may be estimated as
\begin{equation}
\label{RotonPosition}
p_r \approx\sqrt{- \dfrac{2 m \nu }{\beta \hbar^2\ln(2 m R_{\uparrow\downarrow}^2 \nu/ \beta \hbar^2)}},
\end{equation}
and one may readily verify that $p_r\sim \sqrt{n_c^{(2)}} \ll R_{\uparrow\downarrow}^{-1}$. The ensuing tendency of the system to separate into alternating immiscible domains would compete with the pairing and may preclude formation of the $p$-wave spinor molecular superfluid, which is known to be a stable ground state of the system at $\nu<0$ \cite{Radzihovsky2009}. An intriguing outcome of the two competing mechanisms will be explored in our future work.

To close this section, let us discuss the stability of a mixture with respect to possible relaxation into an $s$-wave molecular superfluid. Two general remarks are in order here. First, the $s$-wave bound state underlying a centrifugal $p$-wave scattering resonance in 2D is deep. This is in stark contrast to the 3D setting, where the $p$-wave resonance emerged from a weakly-bound $s$-wave state \cite{Andreev2020}. This seems to be a general trend (see Appendix \ref{RectangularWell}), also recovered within a realistic model considered in Section \ref{Implementation}. Second, in 2D both the elastic $\sigma_e$ and inelastic $\sigma_r$ scattering cross-sections scale generically as $k^{-1}$. One may then expect that inclusion of long-range repulsive forces would suppress the ratio $\sigma_r/\sigma_e$ at low energies.

\section{Physical implementation}

\label{Implementation}

\subsection{The model}

\begin{figure}[t]
\includegraphics[width=0.9\columnwidth]{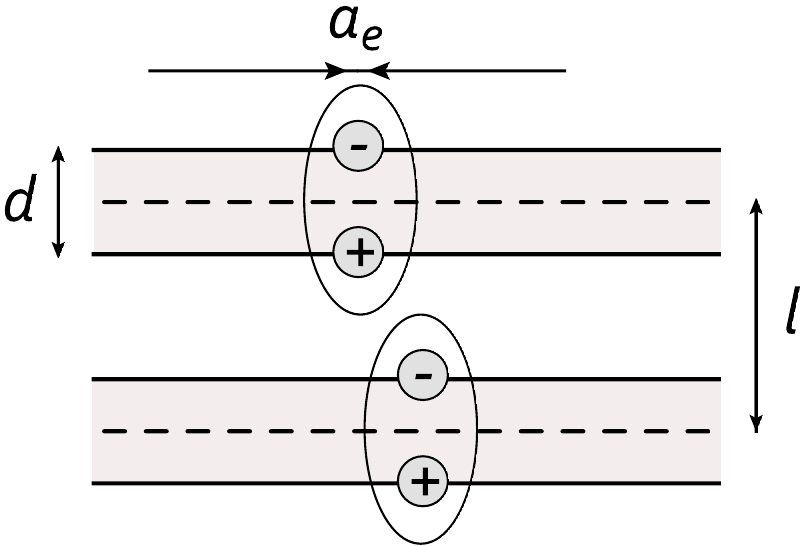}
        \caption{As a particular platform where magnetoroton can be observed experimentally, we propose a double-layer structure with interlayer spacing $l$. Each exciton layer consists of an electron and a hole layer spatially separated by the distance $d$. }
        \label{figLayer}
\end{figure}

Remarkably, in contrast to the 3D case addressed in our previous study \cite{Andreev2020}, in 2D a $p$-wave scattering resonance builds upon a tightly-bound $s$-wave state (see Appendix~\ref{RectangularWell}), which considerably enlarges the choice of potential experimental implementations. Here, we show that the magnetoroton may be observed in a system of dipolar bosons residing in a pair of spatially separated layers, as schematically illustrated in Fig.  \ref{figLayer}. The labels ``$\uparrow$'' and ``$\downarrow$'' may be conveniently ascribed to the top and bottom layers, respectively. The corresponding interaction potentials read
\begin{subequations}
\begin{align}
\label{Repulsive}
V_{\uparrow\uparrow}(r)&=V_{\downarrow\downarrow}(r)=\frac{\hbar^2r_\ast}{m r^3}\\
\label{Attractive}
V_{\uparrow\downarrow}(r)&=\frac{\hbar^2}{m}\frac{r_\ast(r^2-2l^2)}{(r^2+l^2)^{5/2}},
\end{align}
\end{subequations}
where $r_\ast\equiv me^2d^2/\kappa \hbar^2$ is the dipolar length and $l$ is the distance between the layers. The absence of tunneling between the layers ensures statistical distinguishability of thus defined components of a mixture.

The 2D dipolar bosons may be realized with semiconductor excitons formed of spatially separated electrons and holes. Hence, each boson layer $\sigma=\{\uparrow,\downarrow\}$ defined above would correspond to an electron-hole bilayer. Such heterostructures have already been implemented on the basis of epitaxial quantum wells (QWs) and interlayer dipolar attraction has been experimentally demonstrated \cite{Hubert2019, Butov2021}. As we shall deduce below, the magnetoroton formation would require the ratio
\begin{equation}
g\equiv \frac{r_\ast}{l}=\frac{d}{a_e}\times\frac{d}{l}\sim 10,
\end{equation}
which, by virtue of the constraint $d\lesssim l$, may only be achieved at $d\gg a_e$. Here $a_e=\hbar^2\kappa/m_e e^2$ is the electron Bohr radius with $\kappa$ being the dielectric constant of the surrounding material. On the one hand, the condition $d\gg a_e$ excludes binding of excitons within each layer due to the electron-hole exchange \cite{Andreev2015}, thus ensuring implementation of the purely repulsive potential \eqref{Repulsive}. On the other hand, the electrons and holes should not be too far apart in order to avoid dissociation of excitons. In view of that latter constraint, we believe that a more suitable setting would be a Van-der-Waals heterostructure consisting of atomically thin layers of transition metal dichalcogenides (TMDs), say, MoS$_2$. The stability of dipolar excitons in TMDs has been addressed in Ref. \cite{Fogler2014}. At  $d/l\lesssim 1$ one would need $d/a_e\sim 10$, which corresponds to the Mott dissociation densities $n_M a_e^2\sim 10^{-4}$. Note, that recombination of the electrons and holes in the nearby layers separated by the distance $l-d$ may be suppressed by application of a transverse electric field.

\begin{figure}[t]
\includegraphics[width=0.9\columnwidth]{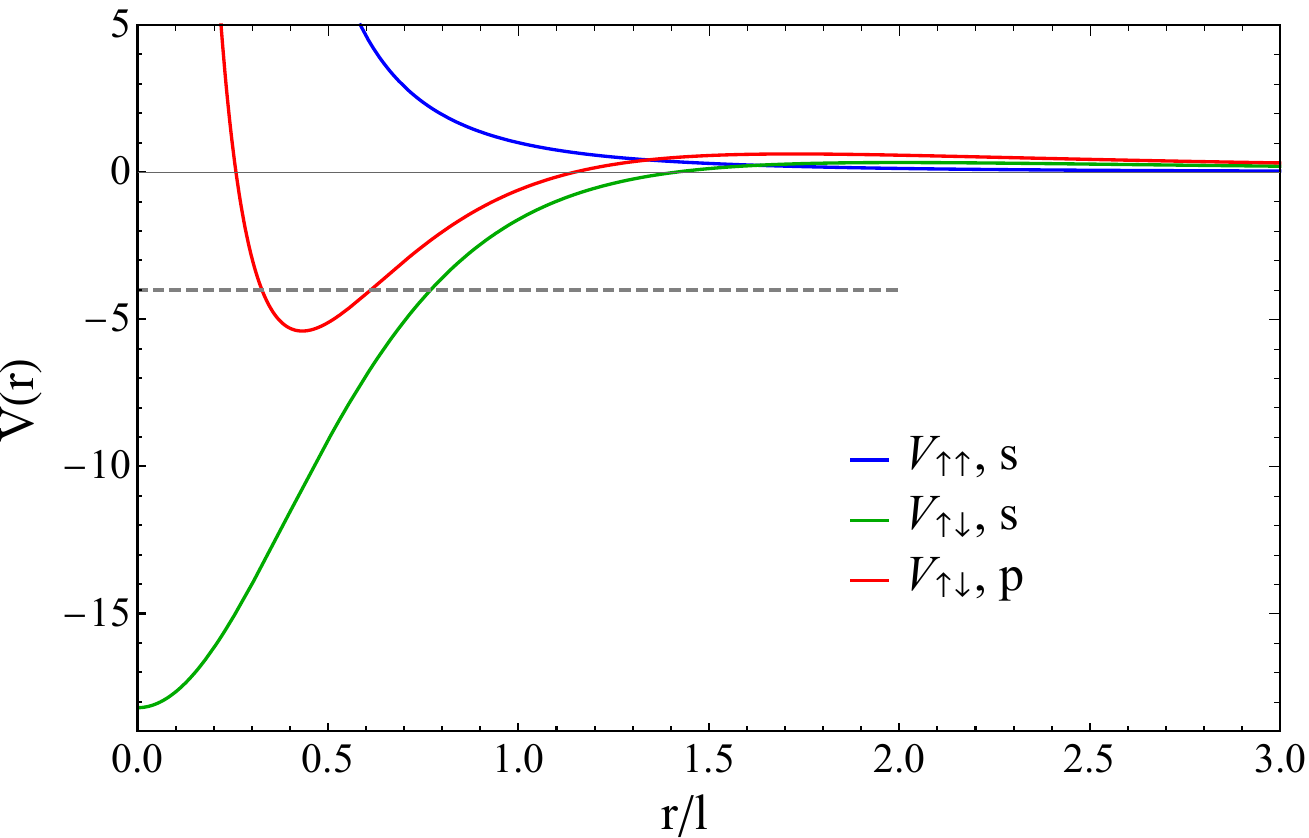}
        \caption{Two-body scattering potentials in various channels in the units of $\hbar^2/ m l^2$ for $g=9.1$ ($p$-wave resonance regime). Orbital momentum contribution $l^2/r^2$ is taken into account for the effective $p$-wave potential $V_{\uparrow \downarrow}, \, p$. The dashed gray line shows the energy of tightly bound $s$-wave state in the interlayer channel with the energy $\approx -4 \hbar^2/ m l^2$. }
        \label{figV}
\end{figure}

The ultra-cold limit for this model is achieved at $k r_\ast\ll 1$. The anomalous scattering corrections $g^\ast_{\sigma\sigma^\prime}(k)=-2\pi\hbar^2/m kr_\ast$ \cite{Shlyapnikov2011, Baranov2011} are identical for all channels and do not contribute to the magnon dispersion (see Section \ref{Generalities}). The dipolar length $r_\ast$ sets the scattering lengths in the intra-layer channels: $a_{\uparrow\uparrow}=a_{\downarrow\downarrow}\sim r_\ast$. The inter-layer channel contains at least one bound state for all $g$ \cite{Klawunn2010}. At $g\ll 1$ this would be a (single) weakly-bound state so that one would have $a_{\uparrow\downarrow}\ll r_\ast$ and the mixture would be far from the conventional immiscibility boundary defined by $a_{\uparrow\downarrow}=a_{\uparrow\uparrow}=a_{\downarrow\downarrow}$ (Section \ref{Generalities}). On a short timescale defined by the emission of the lattice phonons, an $s$-wave superfluid of interlayer excitonic molecules would form (see Refs. \cite{Andreev2015, Andreev2016} for the mean-field study of $s$-wave exciton pairing). In the opposite limit $g\gg 1$, pertinent to our case, one would have $a_{\uparrow\downarrow}\sim r_\ast$, and the conventional miscibility criterion $a_{\uparrow\downarrow}<a_{\uparrow\uparrow}$ should be verified by an explicit calculation. The interlayer potential in this case may support several bound states, with at least one being tightly bound \cite{Baranov2011}. That latter limit is of interest to our study.

\subsection{Results and discussion}

In order to study scattering in realistic potentials given by Eqs.~\eqref{Repulsive} and~\eqref{Attractive}, we perform numerical simulations of the 2D Schr\"{o}dinger equation for the radial part of the wave function of the relative particle motion $\psi^{(+)}_{m,k}(r)$ ($m=0$ corresponds to $s$-waves, $m=1$ -- to $p$-waves). We assume that using the same notation for the particle mass $m$ should not cause any confusion, since the latter would always enter as a combination $\hbar^2/m$. After some transformations, the radial Schr\"{o}dinger equation may be recast in the form
\be \label{Schrod}
  \frac{1}{\rho} \partial_\rho  \left(\rho \partial_\rho \psi^{(+)}_{m,k}\right) - \left[ \frac{m^2}{\rho^2} - (k l)^2 \right]\psi^{(+)}_{m,k} - \frac{m l^2 V(\rho)}{\hbar^2}\psi^{(+)}_{m,k} =0. \nn \\
\ee
where $\rho=r/l$. Note that eigenfunctions with $k^2<0$ correspond to the bound states. The scattering states $\psi^{(+)}_{m,k}(r)$ correspond to $k^2>0$. The scattering  states may be used to obtain the scattering phase shifts $\delta_m(k)$~\cite{LandauQM}. We fit the wave functions far from the scattering potential (at $r \gg l$ or, equivalently, $\rho \gg 1$) with their asymptotic
\be \label{Phase}
  \psi^{(+)}_{m,k}(r) \propto \sqrt{\frac{2}{\pi k r}} \cos{\left[k r - \frac{m \pi}{2} - \frac{\pi}{4} + \delta_m(k) \right]}
\ee
and obtain the corresponding partial-wave amplitudes by using the identity
\begin{equation}
\label{PartialWaves}
f_m(k)=\frac{1}{i}(e^{2i\delta_m(k)}-1).
\end{equation}

\begin{figure}[t]
\includegraphics[width=0.9\columnwidth]{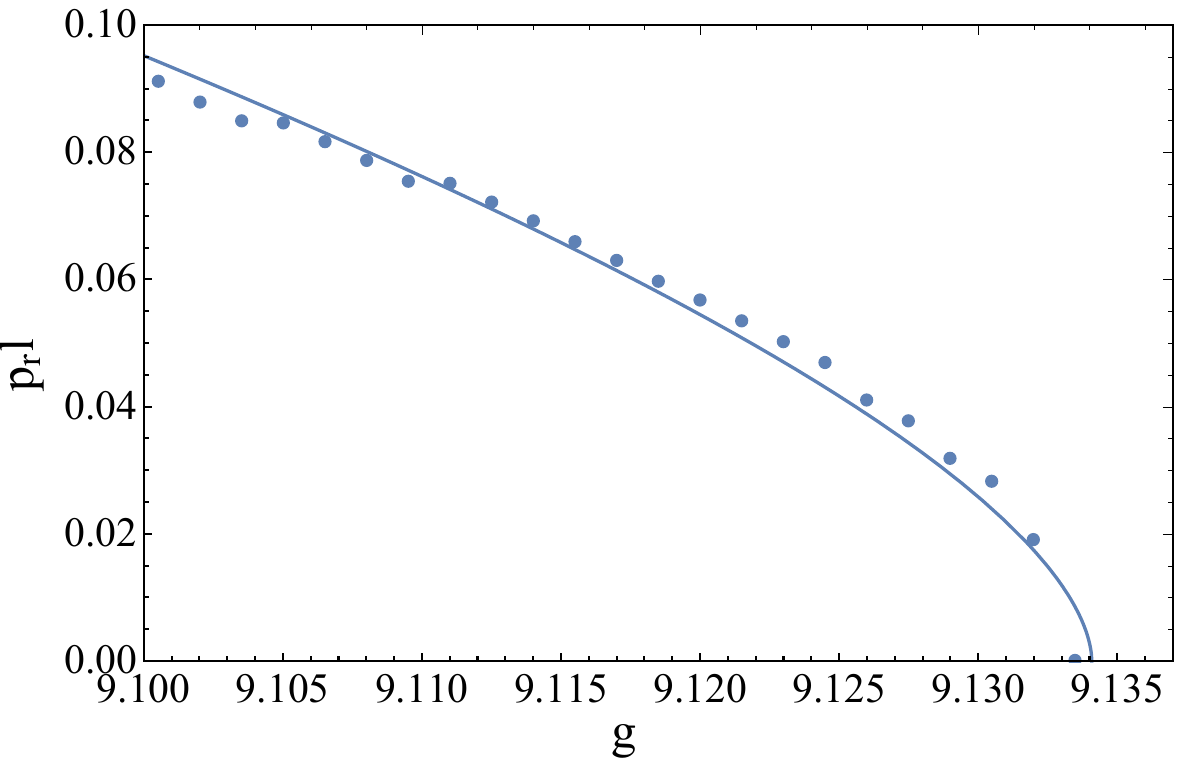}
        \caption{Position of the $p$-wave resonance $p_r$ [maximum of $f_{1,\uparrow\downarrow}(k)$] as a function of the parameter $g$ obtained numerically (dots) and its best fit with the law \mbox{$p_r = l^{-1} \sqrt{\textrm{const} (g^{(c)}-g)/ \ln{(g^{(c)} - g)}}$} (line). }
        \label{figP0}
\end{figure}

\begin{figure}[t]
\includegraphics[width=0.9\columnwidth]{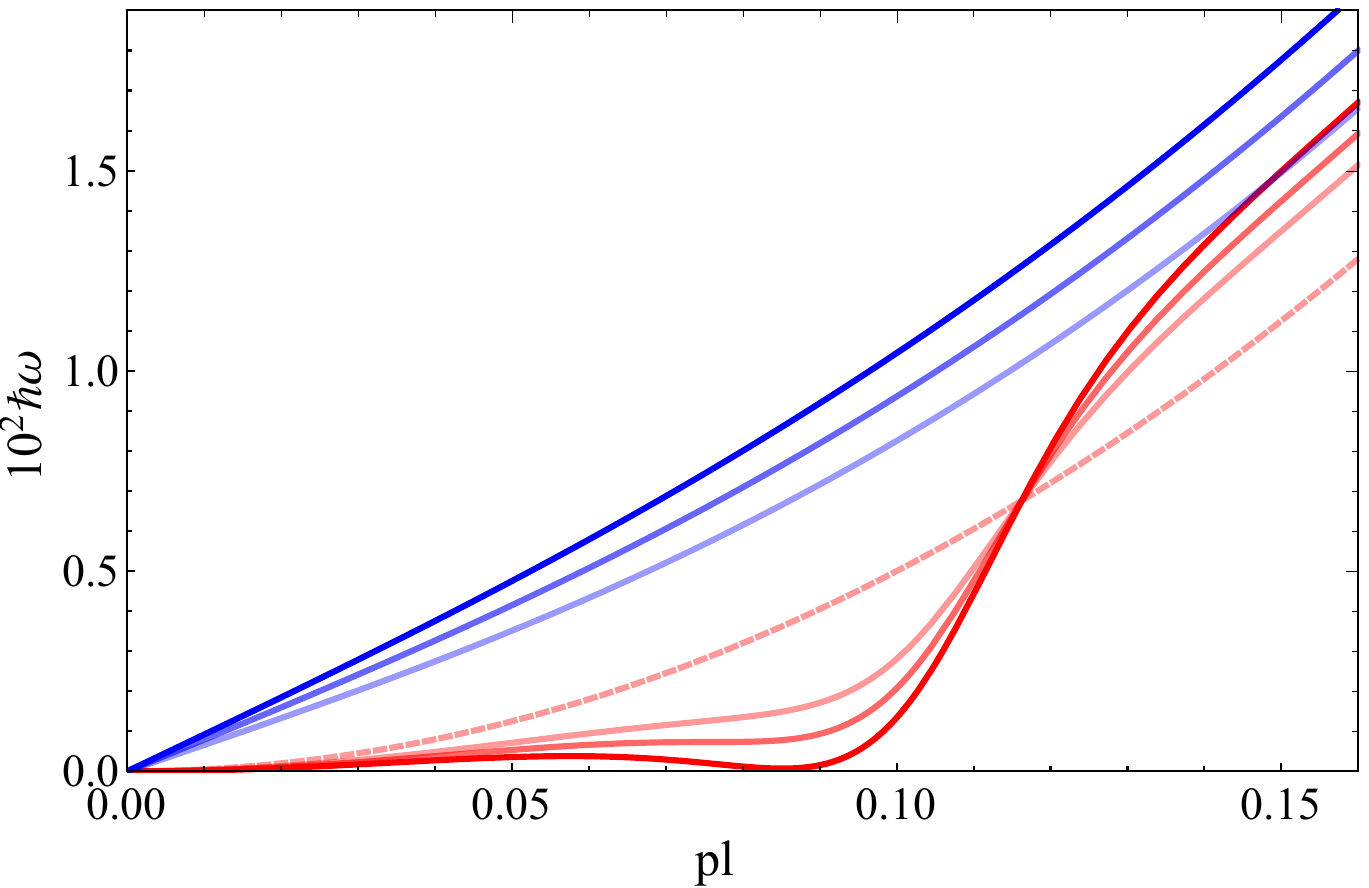}
        \caption{The elementary excitation spectrum of a binary mixture of dipolar bosons in a bilayer structure shown in Fig. \ref{figLayer}. The inter-layer distance $l$ has been used to define the units of energy, momentum and density. The increasing opacity of the curves corresponds to increasing density: the maximal opacity corresponds to $ n l^2 = 0.001 $, intermediate to $n l^2 = 0.0008$, and minimal to $n l^2 = 0.0006$. The magnon and phonon branches are shown using red and blue colors, respectively. The value of detuning is fixed at $g^{(c)} - g = 0.025$, which corresponds to the immediate vicinity of the inter-layer $p$-wave resonance. Dashed line shows the magnon spectrum away from the $p$-wave resonance. Upon increasing the density $n$, the magnon branch develops a minimum which eventually turns into an instability. The phonon branch slightly increases its slope.  }
        \label{figRoton}
\end{figure}

After subtracting the anomalous scattering contributions, we observe a behavior similar to the rectangular-well model (Appendix \ref{RectangularWell}). The $p$-wave resonant scattering is realized at $g \sim 10$. To be more precise, $g^{(c)} \approx 9.134$ and the position of the $p$-wave resonance is given by (see Fig.~\ref{figP0})
\be \label{p0real}
  p_r \approx \frac{1}{l} \sqrt{- 0.9 \frac{g^{(c)} - g}{ \ln{(g^{(c)} - g)}}},
\ee
as obtained by fitting the numerical results with the generic Eq. \eqref{RotonPosition}. Fig.~\ref{figP0} shows that it works quite well. Importantly, in this regime, the interlayer coupling results in a unique tightly bound $s$-wave state with energy $\varepsilon_s \approx -4 \hbar^2/ m l^2$. The radius of the $s$-wave molecule may be estimated as $a_s=\hbar^2/\sqrt{m\varepsilon_s}\approx l/2$. The corresponding $s$-wave scattering lengths are $a_{\uparrow \uparrow} \approx 3.85 l$ and $a_{\uparrow \downarrow} \approx 3.45 l$, and thus satisfy the conventional miscibility criterion $a_{\uparrow\downarrow}<a_{\uparrow\uparrow}$.

For quantitative analysis of the magnon spectrum, we proceed as follows. One can see that $a_{\uparrow \uparrow} \approx a_{\uparrow \downarrow}$, so we can rewrite Eq.~\eqref{mu2D} as $- x^2_c \ln{x^2_c} = 16 \pi n l^2 \equiv \xi$, where we have defined $x_c = p_c l$. The approximate solution reads $x^2_c = -\frac{\xi}{\ln{\xi}}$ provided that $\ln{\xi^{-1}} \gg 1$. The quantity $x_c$ gives a dimensionless momentum scale where the linear phonon dispersion law turns into the quadratic (free-particle) dependence. For the magnon branch, on the other hand, the corresponding scale is given by $y^2_c=\frac{\xi}{(\ln\xi)^2}$, and one may see that $y_c\ll x_c$. Hence, taking into account that $p_r$ is on the order of $p_c$, one may approximate Eq. \eqref{Magnon3D} by its free-particle form $\hbar\omega(\bm p)\approx\hbar^2 p^2/2m_\ast$, where the effective mass is defined by Eq. \eqref{MagnonMass2D_generic}. Since the $p$-wave scattering amplitude has a sharp maximum equal to $1$ at $p=p_r$, we arrive at the following condition for the magnetoroton to touch zero: $(p_r l)^2 = 2 n^{(2)}_c l^2$. This allows one to express the critical density $n^{(2)}_c$ as a function of the detuning $\nu\propto (g^{(c)} - g)$. In dimensionless units one obtains
\begin{equation*}
   n^{(2)}_c l^2 \sim - \frac{g^{(c)} - g}{ \ln{(g^{(c)} - g)}},
\end{equation*}
in agreement with the generic formula \eqref{SecondDensity}. This should be compared with the characteristic density $n_c^{(0)}$ for the onset of a minimum
\begin{equation*}
   n_c^{(0)} l^2 \sim  \frac{g^{(c)} - g}{ \left[\ln{(g^{(c)} - g)}\right]^2}
\end{equation*}
and the density $n_c^{(1)}$ for divergency of the renormalized mass $m_\ast$ at $\bm p\rightarrow 0$:
\be
  n_c^{(1)} l^2 \sim g^{(c)} - g.
\ee
Hence, at small values of the detuning $g^{(c)} - g\ll 1$ one has the following hierarchy of the characteristic densities $n_c^{(0)}\ll n^{(2)}_c\ll n_c^{(1)}$, so that the magnetoroton precedes self-localization at $\bm p\rightarrow 0$. This contrasts with 3D, where both scenarios have been possible \cite{Andreev2020}.

In Fig.~\ref{figRoton} we show the magnetoroton formation in detail. One can see that a minimum in the magnon branch emerges upon increasing the density. Within the narrow window of densities $n$ where the magnetoroton minimum remains above zero energy, its position $p_r$ in the momentum space remains almost unchanged. With the logarithmic accuracy, $p_r$ can be estimated by using Eq.~\eqref{p0real}. For the particular value $g^{(c)}-g = 0.025$ considered here, Eq.~\eqref{p0real} yields $p_r l \approx 0.078$, in reasonable agreement with $p_r l \approx 0.085$ of Fig.~\ref{figRoton}. At the same time, the phonon branch of the spectrum has a standard shape that interpolates between linear (sound) and quadratic (free-particle) dispersions. The change of regimes corresponds to $p  \sim p_c $, and one finds $p_c l\approx 0.12$ for the chosen parameters.

\begin{figure}[t]
\includegraphics[width=0.9\columnwidth]{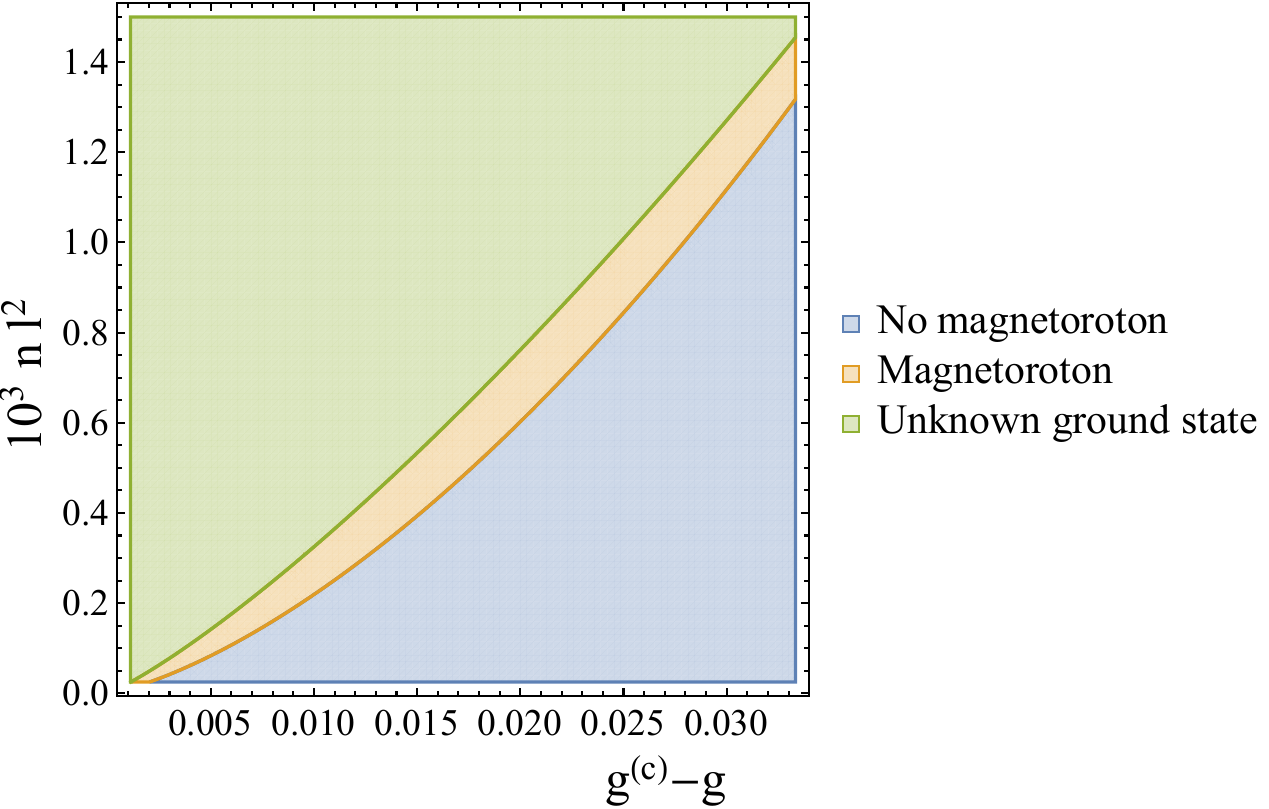}
        \caption{Magnetoroton immiscibility boundary for dipolar bosons in a bilayer structure. The range of "detuning" $\nu\propto g^{(c)} - g$ corresponds to the vicinity of the $p$-wave scattering resonance in the inter-layer interaction of dipoles. The $s$-wave scattering lengths correspond to weak repulsion and satisfy the conventional miscibility criterion $a_{\uparrow\downarrow}<a_{\uparrow\uparrow}$. Upon increasing the density $n$ (expressed here in units of the inter-layer distance $l$) at fixed $\nu$, the magnon dispersion develops a minimum (lower boundary of the orange area) which softens and touches zero energy (upper boundary). At that latter point the mixture becomes dynamically unstable. A possible candidate for a new ground state is a dilute $p$-wave crystal of alternating ``polarization" schematically illustrated in Fig.~\ref{GS}.}
        \label{figPhase}
\end{figure}

In Fig.~\ref{figPhase} we sketch the phase diagram in the density-detuning axes where we distinguish among three regimes: (i) there is no roton minimum in the magnon spectrum (low density and or large detuning), (ii) there is the magnetoroton but the system is stable (moderately larger densities and smaller detunings), and (iii) magnetoroton instability, where near the roton minimum the spectrum formally becomes negative and the system can no longer be considered as a mixture of two interacting gases.


\begin{figure}[t]
\includegraphics[width=0.9\columnwidth]{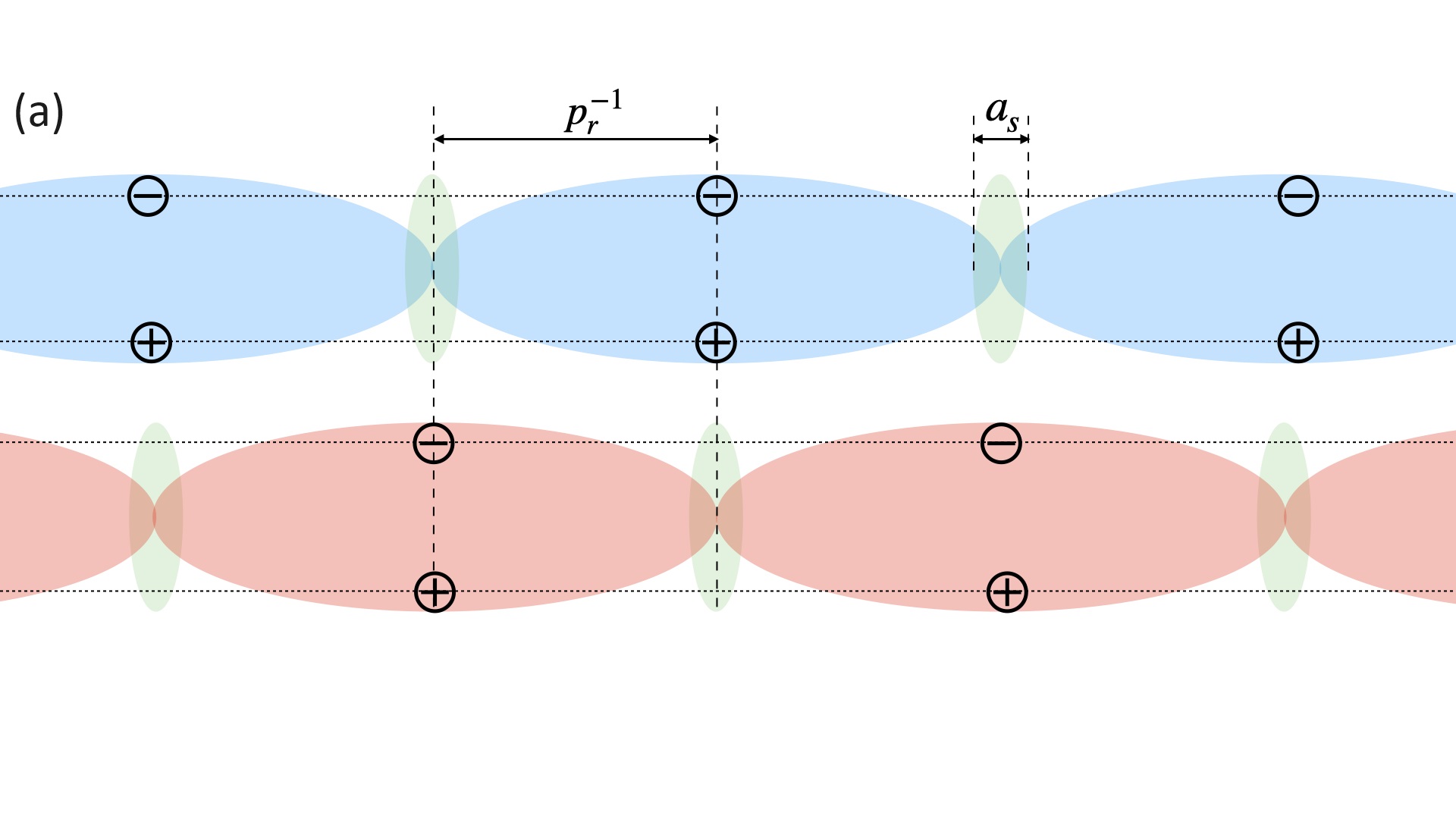}
\includegraphics[width=0.9\columnwidth]{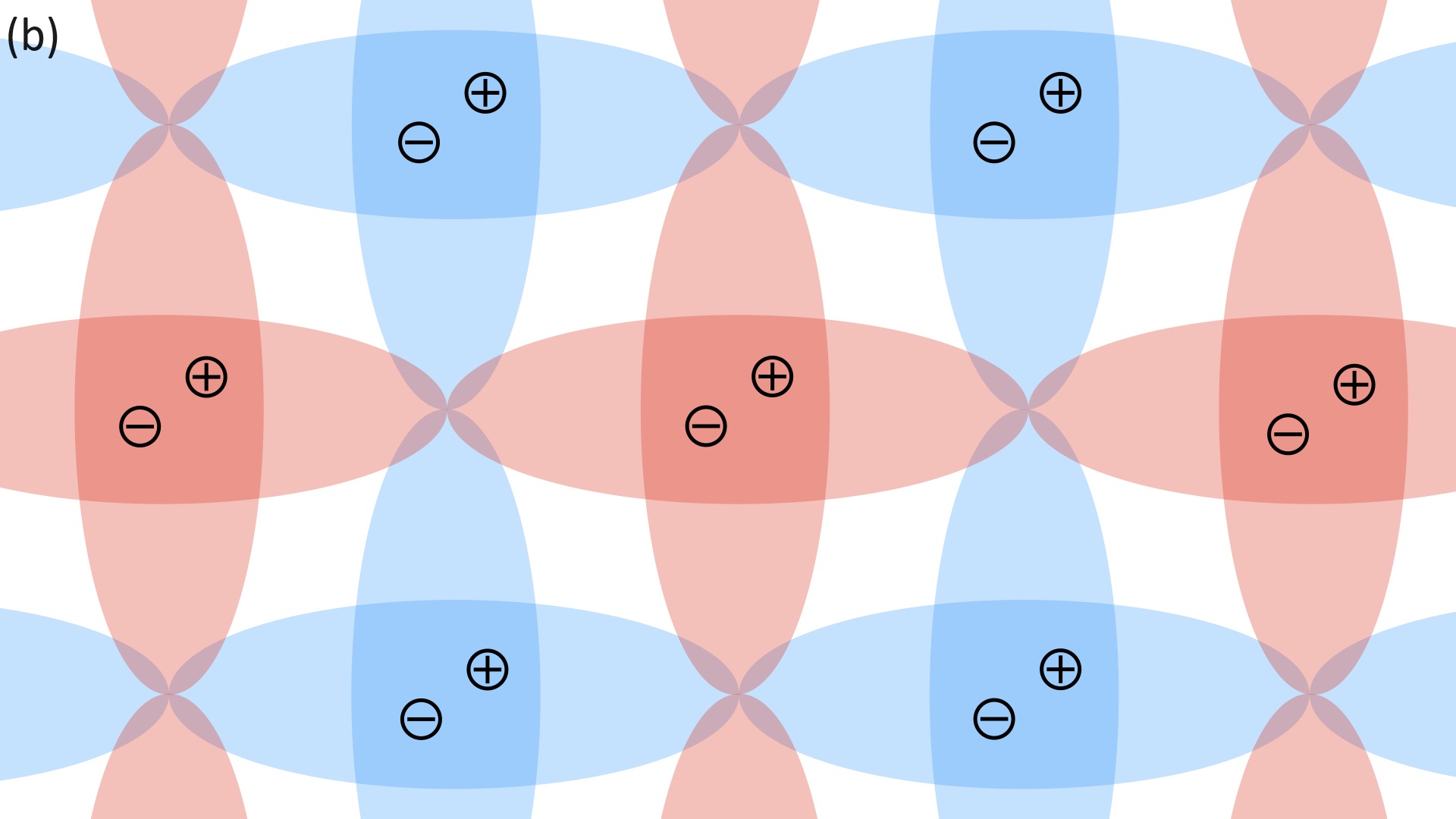}
        \caption{Sketch of a hypothetical new ground state due to the magnetoroton self-localization. Panels (a) and (b) show side-view and top-view, respectively. The dipoles in the upper (``$\uparrow$'') and lower (``$\downarrow$'') layers form $p$-wave orbitals (blue or red, respectively) shifted with respect to each other by the characteristic period on the order of $p_r^{-1}$, thus implementing a stationary phase-separated configuration (a frozen ``polarization'' wave). One has $p_r^{-1}\sim n^{-1/2}\gg R_{\uparrow\downarrow}$. By symmetry, the only possibility for in-plane orbital motion within a unit cell would correspond to a superposition of $m=+1$ and $m=-1$, which suggests a 2D square lattice composed of $p_x$ and $p_y$ molecular orbitals. Note that both the centrifugal barrier and the long-range dipolar repulsion prevent relaxation into deep $s$-wave molecules (green), where an ``$\uparrow$'' dipole would reside on top of the ``$\downarrow$'' dipole in the adjacent layer within the in-plane distance on the order of $a_s=\hbar/\sqrt{m\varepsilon_s}$.  }
        \label{GS}
\end{figure}

Finally, let us comment on the stability of the system with respect to relaxation into an $s$-wave molecular superfluid due to the corresponding tightly bound state of the potential $V_{\uparrow\downarrow}(r)$ (Fig. \ref{figV}). Besides the general arguments provided at the end of Section \ref{Generalities}, one may also try to imagine the new ground state \footnote{As usual for a gaseous BEC, one should rather have in mind a metastable state.} the system would form beyond the magnetoroton instability. A tentative sketch of such a hypothetical state is presented in Fig.~\ref{GS}(a) (side view) and Fig.~\ref{GS}(b) (top view). It reflects two competing mechanisms: the tendency for the components to alternate in the 2D space of the boson translational motion and their $p$-wave pairing. As follows from the above analysis (and from the general consideration of Section \ref{Generalities}), both mechanisms have the same characteristic length scale given by the inverse momentum of the roton $p_r^{-1}$. Our sketch thus represents a resonant $p$-wave crystal, with the nodes in one layer (say, ``$\uparrow$'', blue color) occurring precisely at the antinodes in another layer (``$\downarrow$'', red). Since the node of a $p$-wave orbital (blue or red, respectively) would correspond to the spatial location of an $s$-wave bound state (green), one concludes that in such a dilute crystal the particles would stay far apart from each other. The relaxation into tightly bound $s$-wave states thus would be inhibited both by the centrifugal barrier and the long-range dipolar repulsion. Detailed investigation along these speculative lines will be entertained in our future work.

\section{Conclusions and outlook}

Our consideration shows that a 2D binary Bose-Einstein condensate with weakly repulsive $s$-wave interactions and strong inter-component $p$-wave attraction may display a magnetoroton in the magnon branch of its elementary excitation spectrum. The magnetoroton may be regarded as a form of self-localization of magnons: dramatic enhancement of their mass due to matter-wave refraction. We argue that the condensate suppresses damping of the magnons and stabilizes the magnetoroton close to a $p$-wave scattering resonance. The latter may be obtained from an $s$-wave bound state upon reducing the strength of the inter-component attraction. We notice that in 2D the required depth of the inter-component potential well corresponds to tight binding. This enhances the stability of the system near the $p$-wave resonance and enlarges the choice of candidates for the experimental implementation of our model. As one of such possibilities, we propose a semiconductor heterostructure with a pair of bosonic layers each composed of spatially separated electron and hole layers. Following our general consideration, we have found an inter-species $p$-wave resonance for this particular model and obtained the excitation spectrum featuring a magnetoroton instability upon increasing the total density $n$. The magnetoroton momentum $p_r$ [Eq. \eqref{RotonPosition}] sets the characteristic length scale for two competing mechanisms: in-plane spatial segregation of the components (\textit{i. e.}, formation of a frozen ``polarization" wave) and their $p$-wave pairing. Given that $p_r\sim \sqrt{n}$, we suggest that a possible outcome of such competition may be a resonant $p$-wave crystal, tentatively sketched in Fig. \ref{GS}(a) and (b). Detailed investigation of this intriguing hypothesis will be carried out in our future work.

\begin{acknowledgements}
O. I. Utesov acknowledges financial support from the Institute for Basic Science (IBS) in the Republic of Korea through Project No. IBS-R024-D1.
\end{acknowledgements}

\appendix
\section{Rectangular well model}
\label{RectangularWell}

\begin{figure}[t]
\includegraphics[width=0.9\columnwidth]{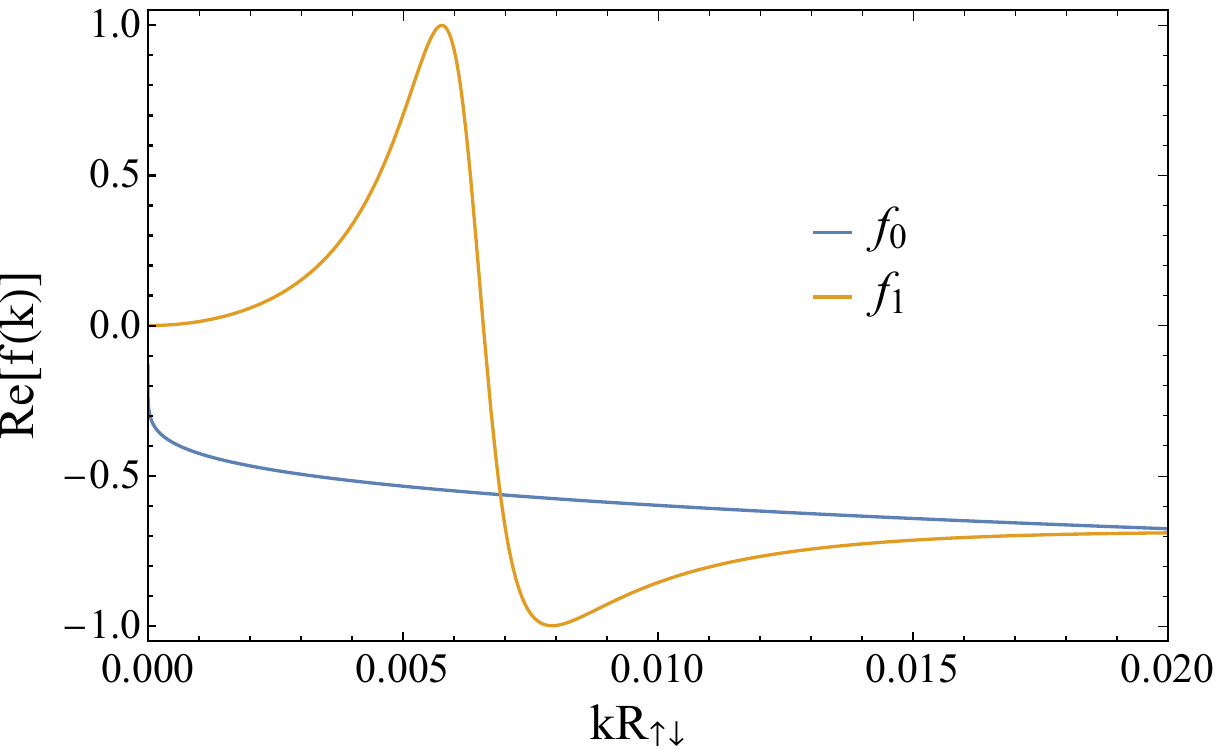}
        \caption{Scattering amplitudes in s- and p- channels for 2D attractive step-like potential in a nearly-resonant case with the detuning parameter $\nu = 0.0002 / m R^2_{\uparrow \downarrow}$. }
        \label{figAmpl}
\end{figure}

In order to develop a feel for the behavior of the magnon spectrum close to a $p$-wave resonance, let us consider a toy model of two-body scattering in a rectangular potential well of depth $U_{\uparrow\downarrow}$ and radius $R_{\uparrow\downarrow}$. The intra-species repulsive interactions $V_{\uparrow\uparrow}(r)=V_{\downarrow\downarrow}(r)$ are modelled by the step-like potentials of equal heights $U_{\uparrow\uparrow}=U_{\downarrow\downarrow}\equiv U$  and radii $R_{\uparrow\uparrow}=R_{\downarrow\downarrow}\equiv R$. The $s$- and $p$-wave scattering amplitudes take the forms of Eq. \eqref{Swave} and Eq. \eqref{PwaveAmplitude}, respectively, with the following parameters:
\be
a_{\uparrow \uparrow} &=& \dfrac{R_{\uparrow \uparrow} \exp{\left(\gamma - \dfrac{I_0(R_{\uparrow \uparrow}\sqrt{m U})}{R_{\uparrow \uparrow}\sqrt{m U} I_1(R_{\uparrow \uparrow}\sqrt{m U})}\right)}}{2}, \\
a_{\uparrow \downarrow} &=& \dfrac{R_{\uparrow \downarrow} e^\gamma}{2}, \\
\beta &=& 1 / \ln{(2 e^{1/2 -\gamma})}, \quad \nu= \beta \frac{U_{\uparrow\downarrow}^{(c)}-U_{\uparrow\downarrow}}{2}, \nn \\ &&\quad U_{\uparrow\downarrow}^{(c)} \approx \frac{5.76\hbar^2}{m R_{\uparrow\downarrow}^2}. \nn
\ee
Here we assume a nearly resonant regime for the inter-species scattering; $I_{0,1}$ are modified Bessel functions of the first kind. The real parts of $f_{0,\uparrow\downarrow}(k)$ and $f_{1,\uparrow\downarrow}(k)$ and shown in Fig.~\ref{figAmpl}. Interestingly, at $\nu=0$ one gets $f_{0,\uparrow\downarrow}(k)\equiv f_{1,\uparrow\downarrow}(k)$. Although the exact identity is an artefact of the rectangular well model, it reflects the important trend: in 2D a $p$-wave resonance builds upon a tightly bound state in the $s$-wave channel. Similar behavior is reproduced within a more realistic model discussed in Section \ref{Implementation}. This is in stark contrast to the 3D case, where only a weakly-bound $s$-wave state could be promoted into a $p$-wave resonance by the centrifugal barrier.

As one can see in Fig.~\ref{figAmpl}, the real part of $f_{1,\uparrow\downarrow}(k)$ has a sharp maximum at $k=p_r$ with $f_{1,\uparrow\downarrow}(p_r)=1$, which corresponds to strong $p$-wave attraction between the particles. At small $\nu$ the corresponding momentum $p_r$ is given by Eq.~\eqref{RotonPosition} and the maximum of $f_{1,\uparrow\downarrow}(k)$ gives rise to the formation of a dip in the magnon dispersion if
\be
  \frac{\hbar^2 (2 p_r)^2}{2 m} \sim \frac{4 \hbar^2 f_{1,\uparrow\downarrow}(p_r) n}{m},
\ee
which yields Eq.~\eqref{SecondDensity} for an estimate of the critical density $n^{(2)}_c$.

\begin{figure}[t]
\includegraphics[width=0.9\columnwidth]{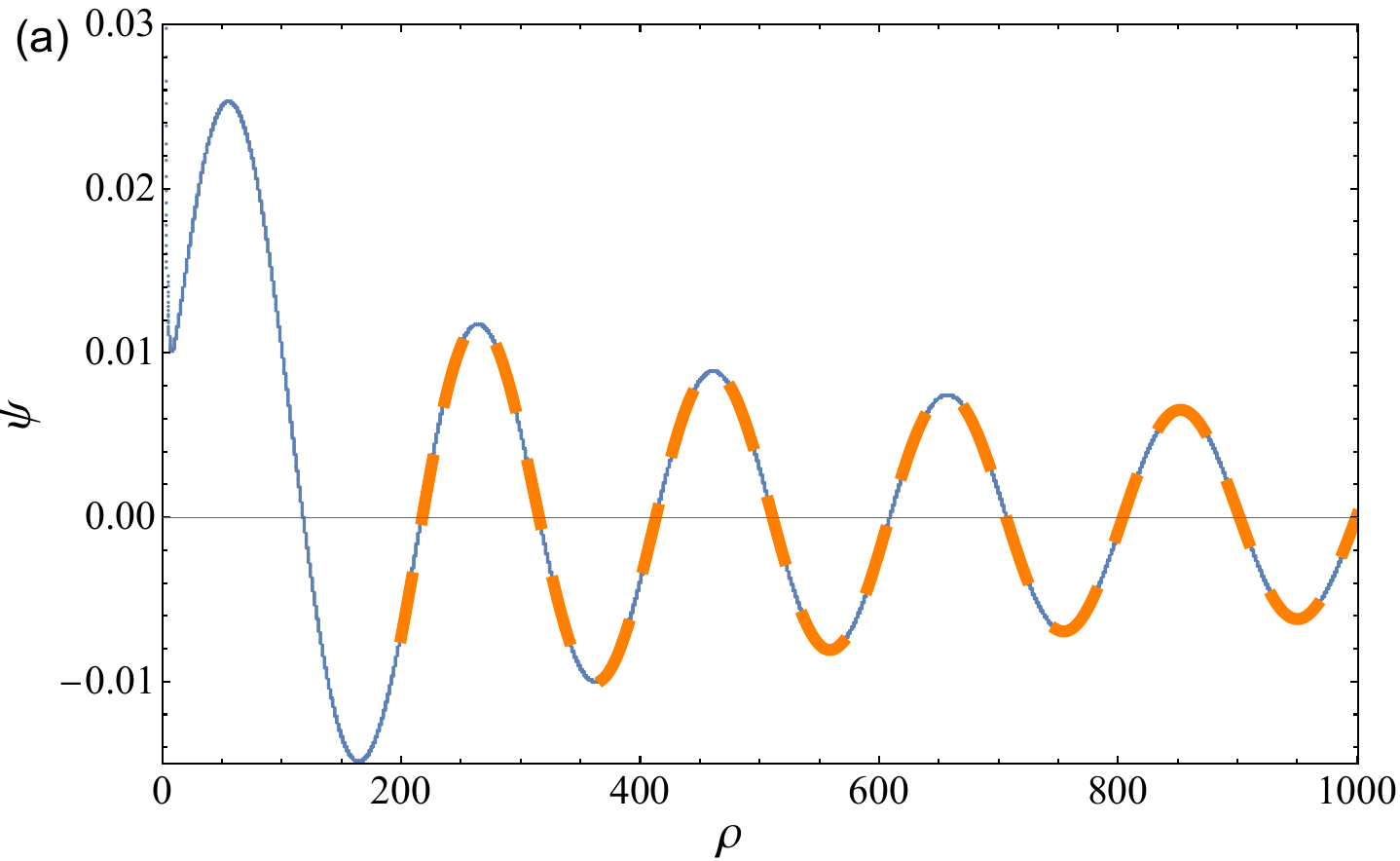}
\includegraphics[width=0.9\columnwidth]{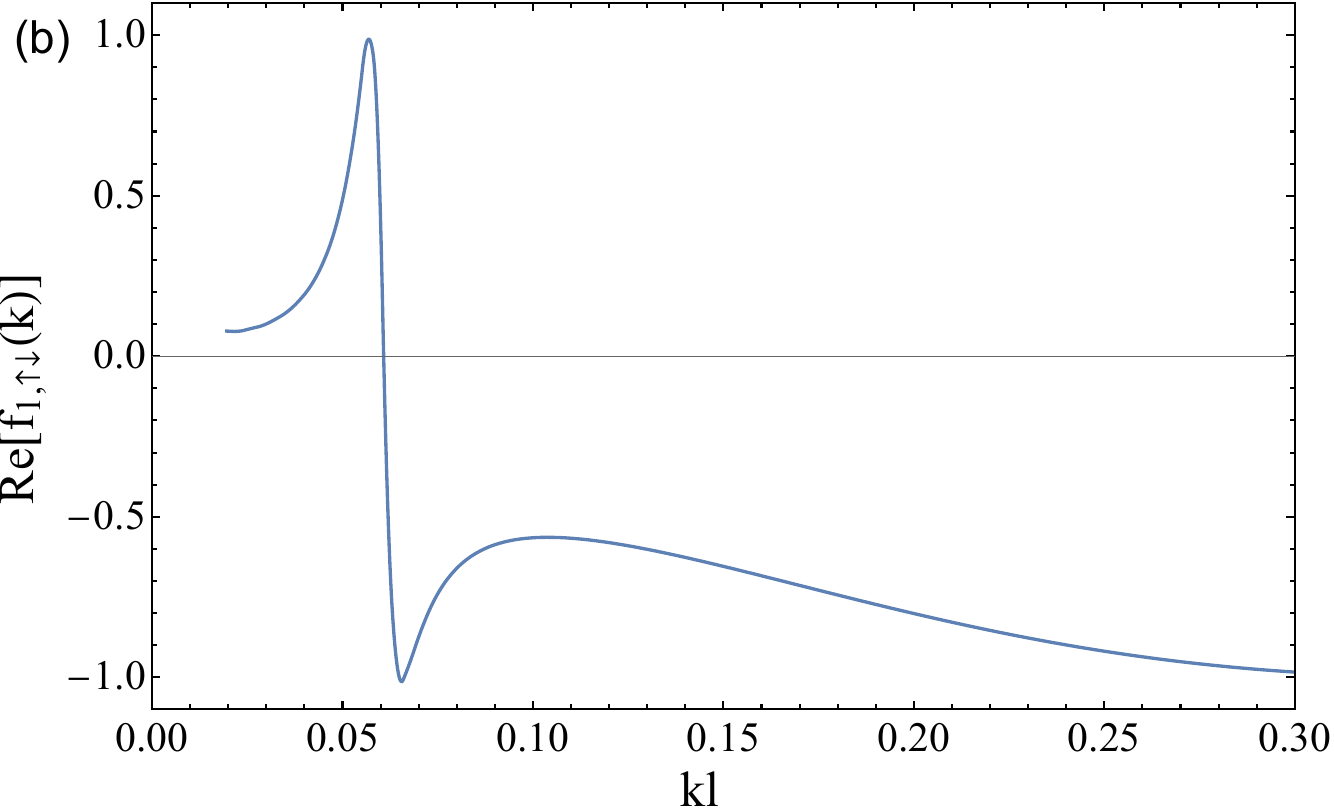}
        \caption{Resonant scattering on the model potential~\eqref{Attractive} in the $p$-wave channel. (a) The wave function corresponding to $kl \approx 0.032$ (10000 blue dots, which look like a curve) and its fit with the asymptotic form~\eqref{Phase} for $\rho \geq 200$ (dashed orange curve). The latter yields the phase $\delta \approx 0.07$.  (b) Dependence of the real part of scattering amplitude $f_{1, \uparrow \downarrow}$ on momentum shows similar to the step-like potential behavior. The parameter $g=9.12$ in both panels. }
        \label{figPReal}
\end{figure}

\section{Details of numerics}
\label{Numerics}

For real experimental potential shown in Fig.~\ref{figV}, we solve the Schrodinger equation~\eqref{Schrod} numerically on a 1D lattice with $N=10000$ or $N=20000$ sites using the standard finite difference scheme. Then, using the developed intuition for step-like potential, we derive important results for scattering amplitudes in $s$- and $p$-channels.

In more detail, we take a certain size $L$ and discretize Eq~\eqref{Schrod} with the step $\Delta \rho = L/N$. Next, $\rho_j = j \Delta \rho, \, j=1...N$. With the use of
\be
  \partial^2_\rho \psi \Bigr|_{\rho_j} &\rightarrow& \frac{\psi_{j+1} - 2 \psi_j + \psi_{j-1}}{(\Delta \rho)^2}, \\
  \partial_\rho \psi \Bigr|_{\rho_j} &\rightarrow& \frac{\psi_{j+1} - \psi_{j-1}}{\Delta \rho}, \nn
\ee
we obtain a system of equations equivalent to $N \times N$ matrix eigenproblem. Importantly, since anomalous scattering does not play a role in our study, we cut $1/\rho^3$ tails of the potential energy at $\rho = g$. For ``boundary conditions'' we use $\rho_0 = \rho_1$ and $\rho_{N+1} = 0$. Finally, the Arnoldi algorithm is employed to find the low-energy eigenstates.

Scattering phases can be found by fitting numerically obtained eigenfunctions with their known asymptotic form~\eqref{Phase}. A particular example of the fit is shown in Fig.~\ref{figPReal}(a) for interlayer $p$-wave scattering and $g=9.12$. Then, we can get the scattering amplitudes using Eq. \eqref{PartialWaves}. For instance, the $k$-dependence of the $f_{1, \uparrow \downarrow}$ is illustrated in Fig.~\ref{figPReal}(b) [cf. Fig.~\eqref{figAmpl}]. Repeating this procedure for various $g$ parameters yields data for position of $p$-wave resonance used in Fig.~\ref{figP0} and for empirical law~\eqref{p0real}.

In contrast, the $s$-wave scattering is weakly dependent on precise $g$ value in the domain of interest ($g \approx 9$). Using the numerical procedure described above and fit of the obtained amplitude with Eq.~\eqref{Swave}, one can find scattering lengths $a_{\uparrow \uparrow}$ and $a_{\uparrow \downarrow}$.

\bibliography{References}
\end{document}